\documentclass[superscriptaddress,showpacs,amsmath,amssymb,twocolumn,nofootinbib]{revtex4-1}

\usepackage{amsmath, amsthm, amssymb, bm, braket}
\usepackage[dvips]{graphicx}
\usepackage[usenames, dvipsnames]{color}
\usepackage{wrapfig}
\usepackage{fancybox}

\begin{document}
\title{Dependence of two-proton radioactivity on nuclear pairing models}

\author{Tomohiro Oishi}
\email[Present affiliation and e-address: ]{{\it Department of Physics and Astronomy 
``Galileo Galilei'', University of Padova, Italy,} toishi@pd.infn.it} 
\affiliation{Helsinki Institute of Physics, P.O. Box 64, 
FI-00014 University of Helsinki, Helsinki, Finland}
\affiliation{Department of Physics, P.O. Box 35 (YFL), 
University of Jyvaskyla, FI-40014 Jyvaskyla, Finland} 
\affiliation{Research Center for Electron Photon Science, Tohoku University, 1-2-1 Mikamine, Sendai 982-0826, Japan}

\author{Markus Kortelainen}
\affiliation{Department of Physics, P.O. Box 35 (YFL), 
University of Jyvaskyla, FI-40014 Jyvaskyla, Finland}
\affiliation{Helsinki Institute of Physics, P.O. Box 64, 
FI-00014 University of Helsinki, Finland}

\author{Alessandro Pastore}
\affiliation{Department of Physics, University of York, Heslington, York YO10 5DD, United Kingdom}


\renewcommand{\figurename}{FIG.}
\renewcommand{\tablename}{TABLE}

\newcommand{\bi}[1]{\ensuremath{\boldsymbol{#1}}}
\newcommand{\unit}[1]{\ensuremath{\mathrm{#1}}}
\newcommand{\oprt}[1]{\ensuremath{\hat{\mathcal{#1}}}}
\newcommand{\abs}[1]{\ensuremath{\left| #1 \right|}}

\def \beq{\begin{equation}}
\def \eeq{\end{equation}}
\def \beqa{\begin{eqnarray}}
\def \eeqa{\end{eqnarray}}
\def \Schr{Schr\"odinger }
\def \twop{2{\it p}}

\def \bir{\bi{r}}
\def \ubir{\bar{\bi{r}}}
\def \bip{\bi{p}}
\def \ubip{\bar{\bi{r}}}

\begin{abstract}
Sensitivity of two-proton emitting decay to nuclear pairing correlation 
is discussed within a time-dependent three-body model. 
We focus on the $^6$Be nucleus assuming $\alpha + p + p$ configuration, 
and its decay process is described as a time-evolution of 
the three-body resonance state. 
For a proton-proton subsystem, 
a schematic density-dependent contact (SDDC) pairing model is employed. 
From the time-dependent calculation, 
we observed the exponential decay rule of a two-proton emission. 
It is shown that the density dependence 
does not play a major role in determining the decay width, 
which can be controlled only by the asymptotic strength 
of the pairing interaction. 
This asymptotic pairing sensitivity can be understood 
in terms of the dynamics of the wave function driven by the 
three-body Hamiltonian, 
by monitoring the time-dependent density distribution. 
With this simple SDDC pairing model, 
there remains an impossible trinity problem: 
it cannot simultaneously reproduce 
the empirical $Q$ value, decay width, and the nucleon-nucleon scattering length. 
This problem suggests that 
a further sophistication of the theoretical pairing model 
is necessary, utilizing the two-proton radioactivity data 
as the reference quantities. 
\end{abstract}

\maketitle

\section{Introduction} \label{Sec:intro}
Description of nuclear pairing correlation has been a major subject in 
recent nuclear physics. 
For instance, in self-consistent meanfield (SCMF) description of atomic nuclei, 
there have been various approaches in order to take the nuclear pairing correlation 
into account \cite{05BB,13BZ,03Dean_rev,03Bender_rev,96Jacek,09Lesi}. 
These approaches based on the SCMF or the nuclear density functional theory (DFT) 
have provided considerable agreements with the measured binding energy as well as 
its odd-even staggering for bound nuclei. 
Even with these efforts, however, the complete character of the 
nuclear pairing correlation has not been revealed. 
For example, whether the phenomenological pairing interaction should have 
the volume or surface type of the density dependence is still an 
open question \cite{2005Sand,2008Pastore,2013Pastore,2013Grasso}. 
At present, one can find several candidates for the sophisticated 
nuclear pairing model \cite{2005Sand,2008Pastore,2013Pastore,2013Grasso,2015Changizi,2010Pillet}. 
In order to validate these models, one may need reference observables to fit 
not only for bound nuclei. 
Also, it should be emphasized that 
the pairing correlation near and beyond the 
neutron and proton driplines could play a fundamental role to determine 
the limit of existence on 
the nuclear chart \cite{2012Markus_Nature,13Olsen}. 

Recently, it has been expected that the two-proton (\twop) radioactivity may provide 
novel information on the nuclear pairing correlation. 
In the true \twop~emission \cite{60Gold,61Gold,09Gri_677,08Blank,2009Gri_rev,2012Pfu_rev}, 
a pair of protons is emitted simultaneously from the parent nucleus, 
whereas the single-proton emission is forbidden or strongly suppressed 
due to the pairing correlation energy. 
The proton-proton pairing plays an essential role to determine 
the observables, especially the released energy ($Q$ value) 
as well as the \twop-decay width or 
lifetime \cite{2009Gri_rev,2012Pfu_rev,2000Gri_PRL,07Gri_III,12Maru,13Deli,2015Lundmark,14Hagi_2n}. 
The released $Q$ value can be related to 
the proton-proton pairing strength in bound nuclei. 
On the other hand, the \twop-decay width has no correspondence in bound systems, 
whose lifetime is trivially infinite. 
Thus, \twop~decays may provide another lodestar with new experimental input to 
optimize and validate the pairing models. 
Thanks to the experimental developments, 
there has been a considerable accumulation of data for the 
\twop-emitting nuclei \cite{08Blank,2009Gri_rev,2012Pfu_rev}. 
On the other side, however, theoretical studies have not been 
sufficient to clarify the relation between the \twop~radioactivity and 
the pairing correlation. 
Because \twop~emission is a typical many-body problem, 
its elucidation could also provide an universal knowledge on 
the multi-particle quantum phenomena in various domains. 
Those include, {\it e.g.} the quantum entanglement \cite{08Bertulani}, 
BCS-BEC crossover \cite{07Marg,07Hagi_01}, and 
Efimov physics \cite{70Efimov,90Bedaque,14Hiyama}. 

In this article, we discuss how the theoretical characters of pairing models 
are reflected on \twop-decay properties, connected to a specific 
interest in sophisticating those models. 
For this purpose, we employ the time-dependent three-body model \cite{14Oishi}, 
whose simplicity enables us to phenomenologically understand the 
pairing model-dependence of \twop~radioactivity. 
We focus on the \twop~emission from the ground state of the $^{6}$Be nucleus, 
assuming the configuration of two valence protons and a rigid $\alpha$ core. 

The formalism of our model is given in Sec. \ref{Sec:form}. 
In Sec. \ref{Sec:calc}, we present numerical calculations and discussions. 
Finally Sec. \ref{Sec:sum} is devoted to summarize the article. 

\section{Three-body model} \label{Sec:form}
Details of the time-dependent three-body calculations are 
present in Ref.\cite{14Oishi}. 
In this article, we employ this method but with some changes. 
The \twop~decay from $^6$Be is described as a time-evolution of 
the two protons in the spherical mean-field generated by the $\alpha$ core. 
The three-body Hamiltonian is given as \cite{1991BE,1997EBH,2005HS,07Hagi_01,2010Oishi}, 
\begin{equation}
 H_{\rm 3b} = h_1 + h_2 + \frac{\bi{p}_1 \cdot \bi{p}_2}{A_{\rm c} m} + v_{\rm p-p}(\bir_1, \bir_2), 
\end{equation}
where $h_i = \bi{p}_i^2 /2\mu + V_{\rm c-p}(r_i)$ ($i=1,2$) is 
the single particle (s.p.) Hamiltonian between 
the core and the $i$-th proton. 
$\mu \equiv m A_{\rm c} / (A_{\rm c}+1)$ is the reduced s.p. mass with $A_{\rm c}=4$. 

The core-proton potential is given as $V_{\rm c-p}(r) = V_{\rm WS}(r) + V_{\rm Coul}(r)$. 
The woods-Saxon potential is expressed as 
\beq
 V_{\rm WS}(r) = V_0 f(r) + U_{ls} (\bi{l} \cdot \bi{s}) \frac{1}{r} \frac{df(r)}{dr}, \label{eq:cp-WS}
\eeq
with a function, 
\beq
 f(r) = \frac{1}{1 + e^{(r-R_f)/a_f}}, 
\eeq
where $R_f=1.68$ fm and $a_f = 0.615$ fm. 
$V_{\rm Coul}$ describes the Coulomb potential. 
For $V_{\rm WS}(r)$ and $V_{\rm Coul}(r)$, we employ the same parameters as in Ref.\cite{14Oishi}, 
in order to reproduce the resonance energy and width 
in the $(p_{3/2})$-channel of $\alpha-p$ scattering, 
$\epsilon_r=1.96$ MeV and $\Gamma_r \simeq 1.5$ MeV, respectively \cite{88Ajzen}. 
Note that this resonance is attributable to the centrifugal potential barrier \cite{14Oishi}.

The $p-p$ pairing potential is introduced 
as $v_{\rm p-p}=v_{\rm p-p}^{(N)} + e^2/\abs{\bir_1-\bir_2}$. 
Here, we employ a schematic density-dependent contact (SDDC) potential, 
\beqa
 v_{\rm p-p}^{(N)}(\bir_1, \bir_2) &=&  w \left( \abs{(\bir_1+\bir_2)/2} \right) \delta(\bir_1-\bir_2), \nonumber \\
 w(r) &=& w_0 \left[ 1 - \eta f(r) \right],  \label{eq:w_SDDC}
\eeqa
for the nuclear pairing interaction: 
two protons have a contact pairing correlation, whose 
density dependence is schematically approximated as the $\eta f(r)$ term. 
For the sake of generality, the density-dependence is not limited to 
have the same $f(r)$ in Eqs. (\ref{eq:cp-WS}) and (\ref{eq:w_SDDC}). 
In this work, however, we use the same function for simplicity, 
except in Sec. \ref{Sec:ddd}. 
Notice also that $w(r\longrightarrow \infty)=w_0$. 
For intrinsic two-nucleon structures, including the dinucleon correlation, 
similar three-body model calculations with SDDC pairing models have been 
utilized \cite{1991BE,1997EBH,2005HS,07Hagi_01,2010Oishi}, with a 
consistency between other theoretical results \cite{05Mats,12Mats,07Hagi_03,10Kiku}. 

With the contact interaction, it is generally known that one should introduce 
the energy cutoff, $E_{\rm cut}$, in order to avoid the ultra-violet divergence \cite{2008Bulgac}. 
In the present case, the bare strength, $w_0$, can be determined so as to 
reproduce the empirical scattering length of nucleons in vacuum, $a_0=-18.5$ fm, 
consistently to the energy cutoff \cite{1991BE,1997EBH}. 
That is, 
\beq
 w_0 = \frac{4\pi^2 \hbar^2 a_0}{m(\pi-2 a_0 k_{\rm cut})}, 
\eeq
with $k_{\rm cut}=\sqrt{mE_{\rm cut}}/\hbar$. 
The cutoff energy is set as $E_{\rm cut}=40$ MeV similarly as in Ref.\cite{14Oishi}, 
yielding $w_0=-767.398$ MeV$\cdot$fm$^3$.

Total expectation value of $H_{\rm 3b}$, which is conserved during 
the time-evolution, corresponds to the released $Q$ value of the three-body decay. 
That is, 
\beqa
 Q_{\rm 2p} &=& \Braket{\Psi(t)|H_{\rm 3b}|\Psi(t)}, \\
 &=& \Braket{h_1+h_2}_{(t)} + \Delta_{\rm pair}(t) \nonumber, \\
 \Delta_{\rm pair}(t) &=& \Braket{\Psi(t)| \frac{\bi{p}_1 \cdot \bi{p}_2}{A_{\rm c} m} + v_{\rm p-p}(\bir_1, \bir_2) |\Psi(t)}, 
\eeqa
where $\Delta_{\rm pair}$ is the pairing correlation energy (PCE). 
In order to reproduce the empirical $Q$ value, $1371 \pm 5$ keV \cite{88Ajzen}, 
we should employ a density-dependence parameter, $\eta=1.04$ in Eq.(\ref{eq:w_SDDC}). 
Namely, the empirical $Q$ value requires 
almost the surface type of the SDDC pairing interaction, 
which imitates the surface version of the density-dependent 
pairing energy in DFT calculations \cite{2005Sand,2008Pastore,2013Pastore}. 
Note also that PCE approximately corresponds 
to the pairing gap when the system is bound.

We assume the $0^+$ configuration for two protons and the $\alpha$ core, 
consistently with the total spin-parity which is also $0^+$ for the ground state of $^6$Be. 
Thus, the eigenstates of the three-body Hamiltonian, 
satisfying $H_{\rm 3b}\ket{E_N}=E_N\ket{E_N}$, 
can be expanded on the $0^+$-configuration basis: 
\beqa
 && \ket{E_N} = \sum_M U_{NM} \ket{\Phi_M}, \label{eq:expansion} \\
 && \Phi_M (\bir_1, \bir_2) 
 = \oprt{A} [\phi_{n_a ljm}(\bir_1) \otimes \phi_{n_b lj(-m)}(\bir_2)]^{0^+}, \label{eq:basis_0p}
\eeqa
with the simplified notation, $M=(n_a, n_b, l,j)$. 
Here $\oprt{A}$ is the anti-symmetrization operator. 
The expansion coefficients, $U_{NM}$, are determined by 
diagonalizing the Hamiltonian matrix for $H_{\rm 3b}$. 
Each s.p. state satisfies $h_{i} \phi_{nljm}(\bir_i) = \epsilon_{nlj} \phi_{nljm}(\bir_i)$, 
with the radial quantum number $n$, 
the orbital angular momentum $l$, 
the spin-coupled angular momentum $j$ and 
the magnetic quantum number $m$. 
We employ the s.p. states up to the $(h_{11/2})$-channel. 
In order to take into account the Pauli principle, 
we exclude the first $s_{1/2}$ state, which is occupied by the protons in the core nucleus. 
The continuum s.p. states ($\epsilon_{nlj} >0$) 
of $V_{\rm c-p}$ are discretized in the radial box of $R_{\rm box}=80$ fm. 
Thus, continuum eigenstates of $H_{\rm 3b}$ are also discretized. 
As we present in Sec.\ref{Subsec:width}, this radial box is sufficiently large 
to provide a good convergence in terms of the decay width. 
\begin{figure}[tb] \begin{center}
     \includegraphics[width = 0.95\hsize]{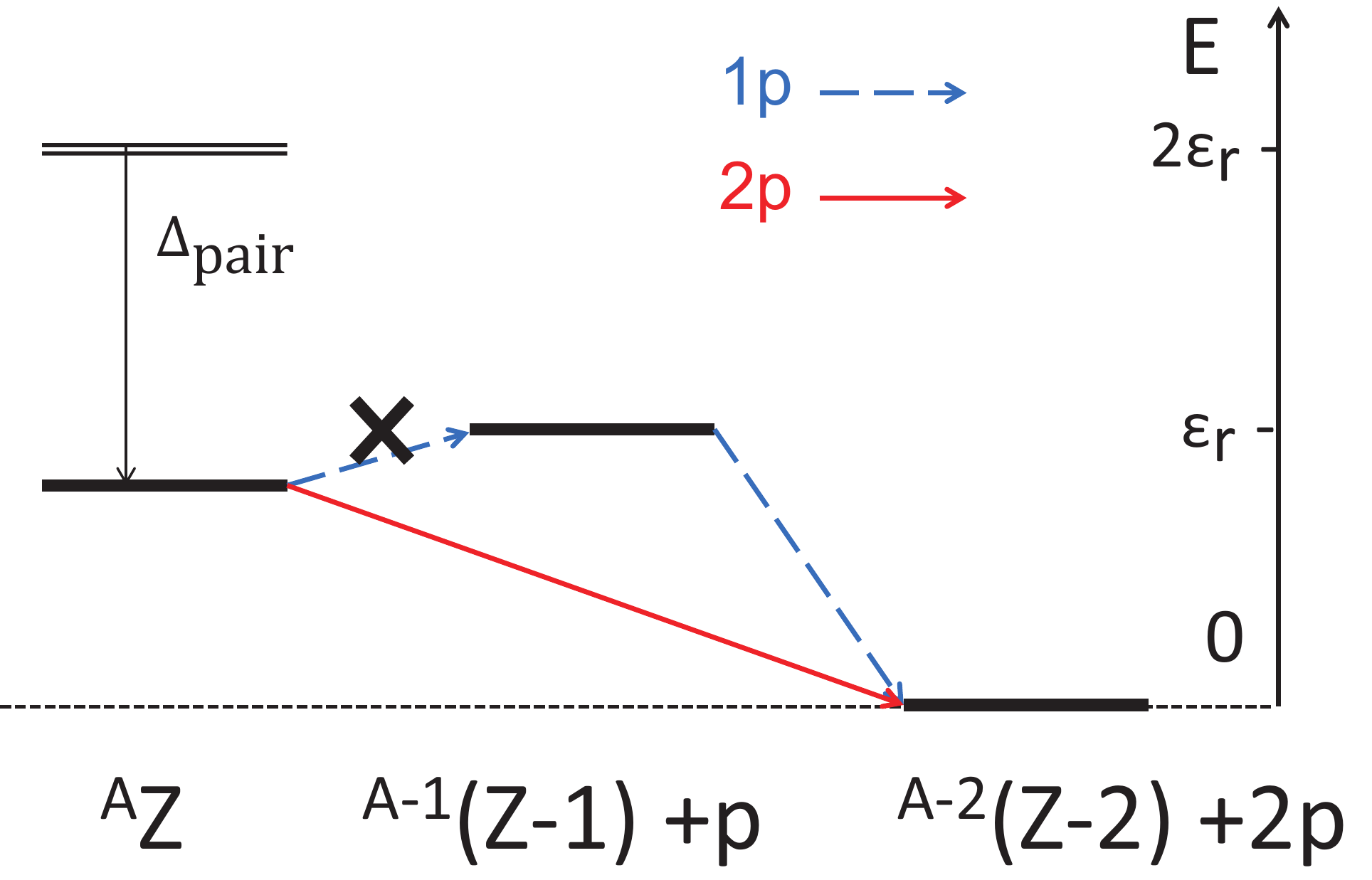}
 \caption{Schematic figure of level scheme, 
in which the correlated \twop~emission becomes dominant. 
} \label{fig:levels}
\end{center} \end{figure}

It is worthwhile to mention that, if one can neglect PCE, 
a \twop-resonance state locates at $Q_{\rm 2p}=\Braket{h_1+h_2}=2\epsilon_r$, 
where $\epsilon_r$ is the $\alpha-p$ resonance energy. 
In this case, where the \twop-wave function keeps the pure $(p_{3/2})^2$ configuration, 
it was confirmed that the decay process is well described as a sequential \twop~emission \cite{14Oishi}. 

Taking PCE into account, 
the resonance energy is decreased mainly due to the attractive pairing interaction. 
Figure \ref{fig:levels} schematically describes this situation. 
In Ref.\cite{14Oishi}, the finite-range, density-independent Minnesota potential 
was employed to describe the pairing force \cite{77Thom}, and then the strongly correlated \twop~emission 
was suggested.\footnote{We found a typo in Table I of Ref.\cite{14Oishi}: ``$S=0~(1)$'' should be corrected as ``$S=1~(0)$''.} 
For the spatial correlation in this process to occur, 
a mixture of other configurations from $(p_{3/2})^2$ plays 
an essential role \cite{2005HS,07Hagi_01}. 
Also, especially with the light core nucleus, 
the induced correlation by the recoil term, 
$(\bi{p}_1 \cdot \bi{p}_2)/A_{\rm c} m$, can be noticeable. 
We check its effect on the $Q$ value in the next section.

\begin{figure}[tb] \begin{center}
     \includegraphics[width = \hsize]{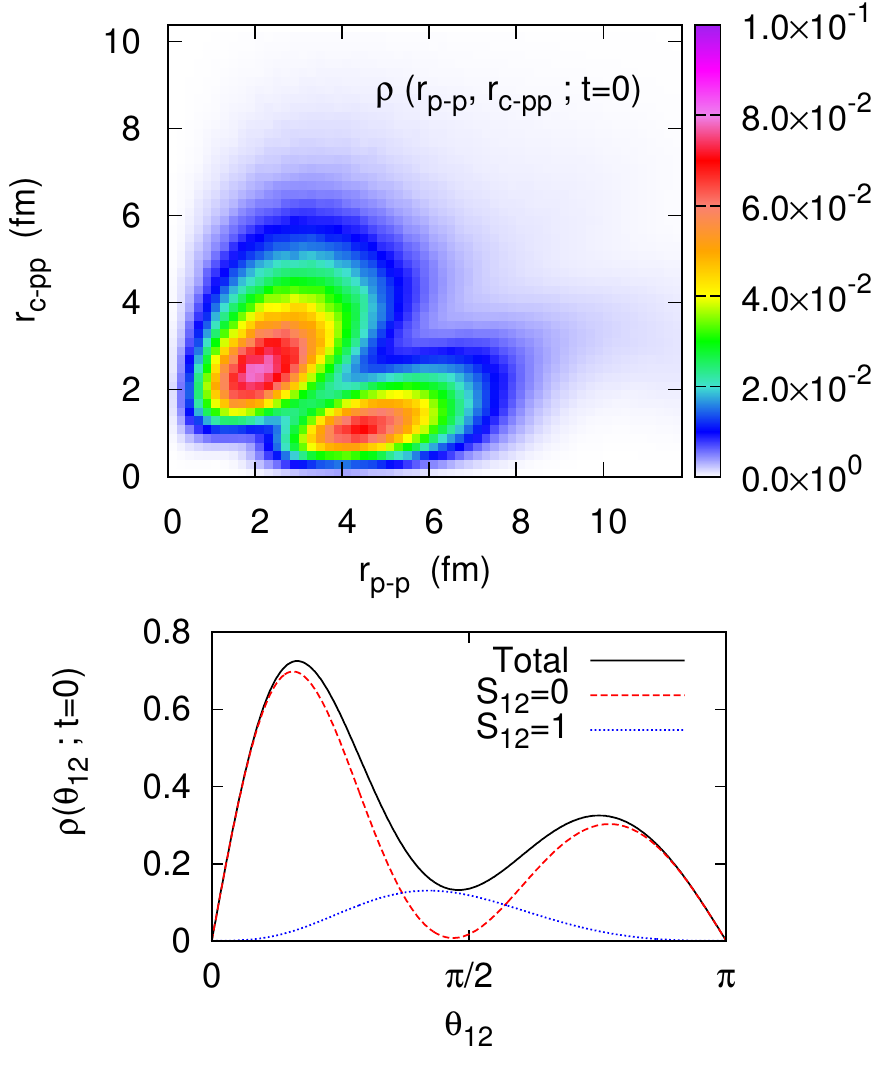}
 \caption{The density-distribution for the initial \twop~state obtained with the surface SDDC pairing interaction. 
(i) Top panel: the distribution as a function of $r_{\rm p-p}$ and $r_{\rm c-pp}$. 
(ii) Bottom panel: 
the angular distribution as a function of the opening angle, $\theta_{12}$, between two protons. 
This is obtained by integrating $\rho(r_1,r_2,\theta_{12})$ for the radial coordinates, $r_1$ and $r_2$. 
The spin-singlet and spin-triplet components are also plotted. 
} \label{fig:1}
\end{center} \end{figure}

\begin{figure*}[tb] \begin{center}
  \includegraphics[width = 0.95\hsize]{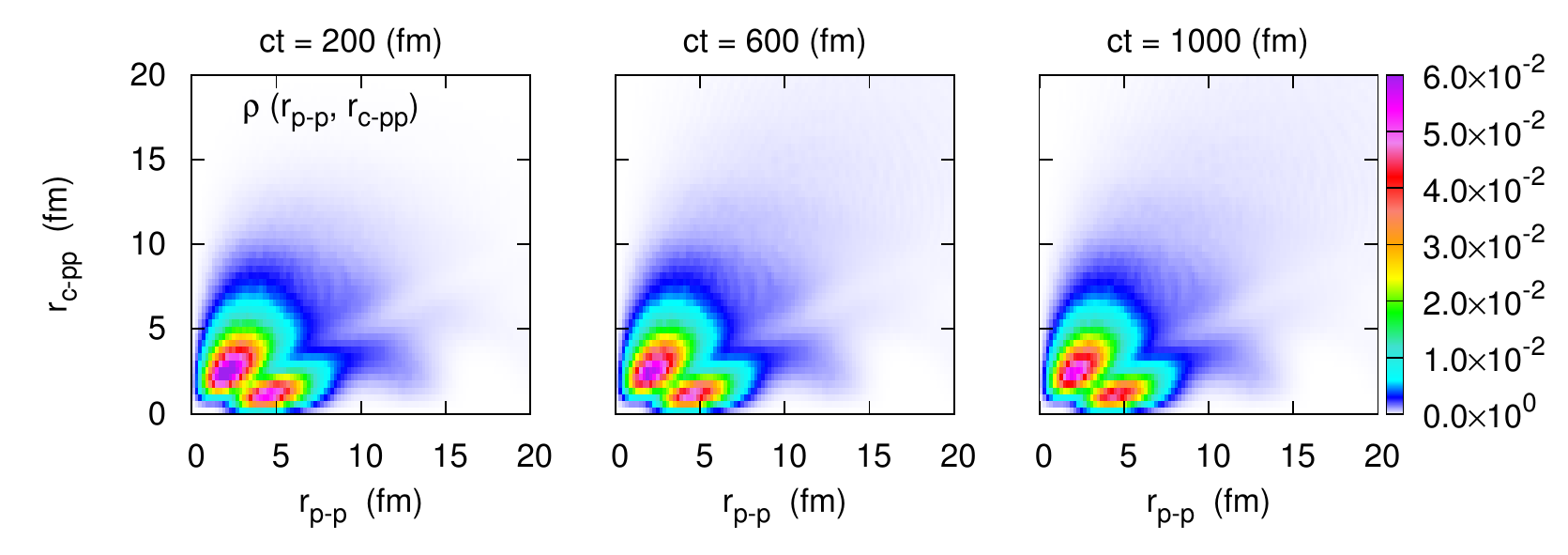}
  \caption{Time-dependent \twop-density distribution plotted as a function of $r_{\rm p-p}$ and $r_{\rm c-pp}$. 
The surface SDDC pairing interaction is employed. 
} \label{fig:Dens_whole_tide}
\end{center} \end{figure*}

\section{Numerical Calculation} \label{Sec:calc}
\subsection{Initial state}
In order to fix the initial state for time-evolution, 
we employ the same confining potential, $V_{\rm c-p}^{\rm conf}(r)$, as in Ref.\cite{14Oishi}. 
This confining potential method has provided good approximations for 
quantum resonance phenomena \cite{87Gurv,88Gurv,04Gurv,12Maru}, together with 
an intuitive way to understand their dynamics. 
The initial state solved within the confining Hamiltonian 
can be expanded on the eigen-basis of the original Hamiltonian: 
$\ket{\Psi (0)} = \sum_N F_N (0) \ket{E_N}$. 
Thus, the time-evolution is represented as 
\beq
 \ket{\Psi (t)} 
 \equiv  \exp \left[ -it \frac{H_{\rm 3b}}{\hbar} \right] \ket{\Psi (0)} = \sum_{N} F_N (t) \ket{E_N}, \label{eq:ex_E} 
\eeq
where $F_N (t) = e^{-itE_N/\hbar} F_N(0)$. 
It is worthwhile to note that the time-invariant discrete energy spectrum can be given as 
\beq
 d(E_N)=\abs{F_N(0)}^2=\abs{F_N(t)}^2. 
\eeq
If one takes the continuous energy limit, $d(E)$ resembles the Breit-Wigner spectrum, 
which characterizes the quantum resonance properties of concerning radioactive process \cite{89Kuku}.

In Fig. \ref{fig:1}, we plotted the normalized density distribution for the initial state. 
That is, 
\beq
 \rho (t;r_1,r_2,\theta_{12}) = 8\pi^2 r_1^2 r_2^2 \sin \theta_{12} \abs{\Psi(t;r_1,r_2,\theta_{12})}^2, 
\eeq
at $ct=0$ fm. 
For convenience, $\rho$ is translated to a function of 
the relative distance between the two protons, $r_{\rm p-p} = (r_1^2+r_2^2-2r_1r_2\cos \theta_{12})^{1/2}$, and 
that between the core and the center of mass of two protons, $r_{\rm c-pp} = (r_1^2 + r_2^2 + 2r_1r_2\cos \theta_{12})^{1/2}/2$. 
From Fig. \ref{fig:1}, we find the similar result in Ref. \cite{14Oishi}, 
where a finite-range Minnesota pairing was used instead: 
the higher peak at $r_{\rm p-p} \simeq 2.0$ fm and $r_{\rm c-pp} \simeq 2.5$ fm, 
as well as at $\theta_{12} \simeq \pi /6$, 
indicates a strong localization of two protons. 
The similar discussion can be found 
in, {\it e.g.} Refs. \cite{1986Kuku, 07Hagi_01}, where the pairing 
correlation as well as the Pauli principle play a fundamental role. 
Notice also that this localization is attributable to the 
spin-singlet configuration, 
suggesting a diproton correlation \cite{2010Oishi}.

\begin{figure*}[tb] \begin{center}
  \includegraphics[width = 0.95\hsize]{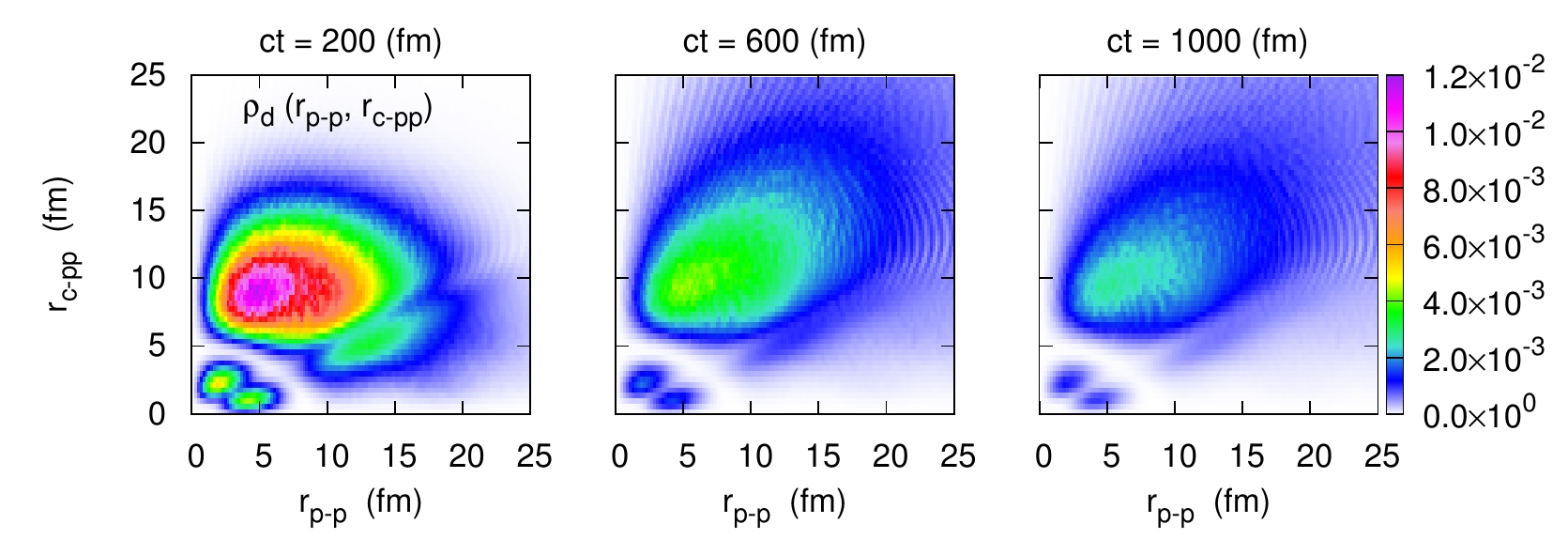}
  \includegraphics[width = 0.95\hsize]{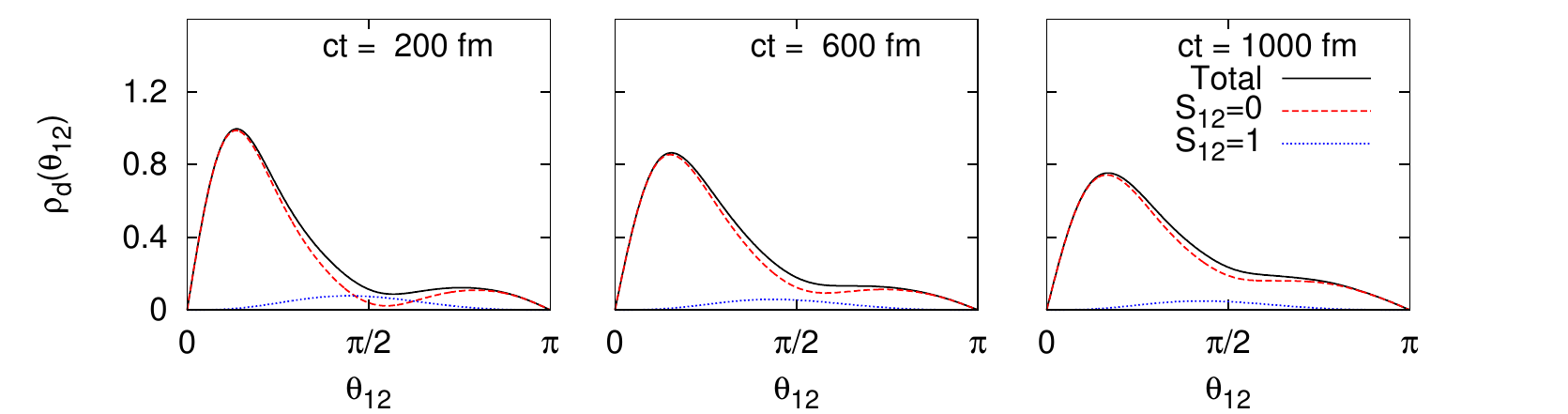}
  \caption{Time-dependent \twop-density distribution of the decay state, $\rho_d(t)$, given by Eq. (\ref{eq:decay_dens}). 
The surface SDDC pairing interaction is employed. 
These are plotted in the same manner as in Fig. \ref{fig:1}. 
} \label{fig:Dens_tide}
\end{center} \end{figure*}

\begin{table*}[tb] \begin{center}
\caption{Parameters for the SDDC pairing interaction used in this article. 
$E_{\rm cut}=40$ MeV. 
Resultant $Q$ values of the \twop~emission of $^6$Be and the corresponding total, 
spin-singlet, and spin-triplet decay widths at $ct=1000$ fm are also present. 
Same results but with the softened, 
finite-range Minnesota pairing model are 
taken from Ref.\cite{14Oishi}. }
\label{table:CP}
  \catcode`? = \active \def?{\phantom{0}} 
  \begingroup \renewcommand{\arraystretch}{1.2}
\begin{tabular*}{\hsize} { @{\extracolsep{\fill}} ccccc cc} \hline \hline
      & $w_0$              & $\eta$ &  $Q_{\rm 2p}$ & $\Gamma$ & $\Gamma_{S=0}$ & $\Gamma_{S=1}$?? \\
      & (MeV$\cdot$fm$^3$) &        &  (keV)       & (keV)    & (keV)         & (keV)??        \\ \hline
~SDDC (this work) & ?$-767.398$   & $1.04$      &  $1370.7$    & $34.7$         & $33.4$         & $1.3$?? \\
~Minnesota \cite{14Oishi} &&        &  $1370.7$   & $88.2$    & $87.1$         & $1.1$?? \\
~Experiment \cite{88Ajzen} &&       &  $1371\pm 5$ & $92\pm6 $ & $-$           & $-$?? \\
\hline \hline
\end{tabular*}
  \endgroup
  \catcode`? = 12 
\end{center} \end{table*}

For this initial state, the $Q$ value is obtained as, 
\beq
 Q_{\rm 2p} = \Braket{\Psi(0)|H_{\rm 3b}|\Psi(0)} = 1.37~~{\rm MeV}
\eeq
with our surface SDDC pairing interaction. 
This includes the negative PCE, where not only the pairing interaction 
but also the induced correlation from the recoil term give finite values. 
That is, 
\beq
 \Delta_{\rm pair} = 
 \Braket{\frac{\bi{p}_1 \cdot \bi{p}_2}{A_{\rm c} m}} + 
 \Braket{v_{\rm p-p}}
 = -6.28~~{\rm MeV}, 
\eeq
where $\Braket{(\bi{p}_1 \cdot \bi{p}_2)/A_{\rm c} m}=-1.46$ MeV 
and $\Braket{v_{\rm p-p}}=-4.82$ MeV. 
Obviously, the pairing interaction makes a major contribution 
in reproducing the empirical $Q$ value. 
In our case, the recoil term effect is also noticeable, 
which exhausts about $25 \%$ of total PCE. 
This feature of the center-of-mass effect may take place 
when the masses of ingredient particles are comparable.

\subsection{Time evolution}
In Fig. \ref{fig:Dens_whole_tide}, we plotted the time-development of \twop~state, in 
terms of the probability-density distribution. 
It is well described that the confined two protons at $ct=0$ are 
released during the time-development. 
The probability-density outside the core-proton barrier gradually increases, 
indicating an evacuation of two protons. 
In order to monitor their decay dynamics more preciously, 
it is helpful to focus on the projected decay state \cite{08Bertulani}. 
That is, 
\beq
 \ket{\Psi_d(t)} \equiv \ket{\Psi (t)} - \beta(t) \cdot \ket{\Psi (0)}, 
\eeq
with $\beta(t) = \Braket{\Psi (0)|\Psi (t)}$. 
Because the initial state is well confined, this projected decay state mainly 
represents the outgoing components released from around the core. 
In Fig. \ref{fig:Dens_tide}, 
we plot the density distribution of the projected decay state 
normalized at each point of time. 
That is, 
\beq
 \rho_d(t;r_1,r_2,\theta_{12}) = \frac{8\pi^2 r_1^2 r_2^2 \sin \theta_{12} \abs{\Psi_d(t;r_1,r_2,\theta_{12})}^2}{N_d(t)}, \label{eq:decay_dens}
\eeq
where $N_d(t)=\Braket{\Psi_d(t) | \Psi_d(t)}=1-\abs{\beta(t)}^2$ is the decay probability. 
In Fig. \ref{fig:Dens_tide}, 
the strongly correlated \twop-emission is suggested with our surface SDDC pairing model. 
The diproton correlation, which can be detected as a peak at 
$r_{\rm p-p} \simeq 5$ fm and $r_{\rm c-pp} \simeq 10$ fm, 
as well as at $\theta_{12} \simeq \pi /6$, 
is dominant during the time-evolution. 
Notice also that the sequential \twop~emission, which is graphically indicated as a ridge 
along the $r_{\rm c-pp} \simeq r_{\rm p-p}/2$ line \cite{14Oishi}, is strongly suppressed. 
This dynamical behavior of protons is similar to that suggested 
from the finite-range Minnesota pairing model \cite{14Oishi}.

\begin{figure}[tb] \begin{center}
  \includegraphics[width = \hsize]{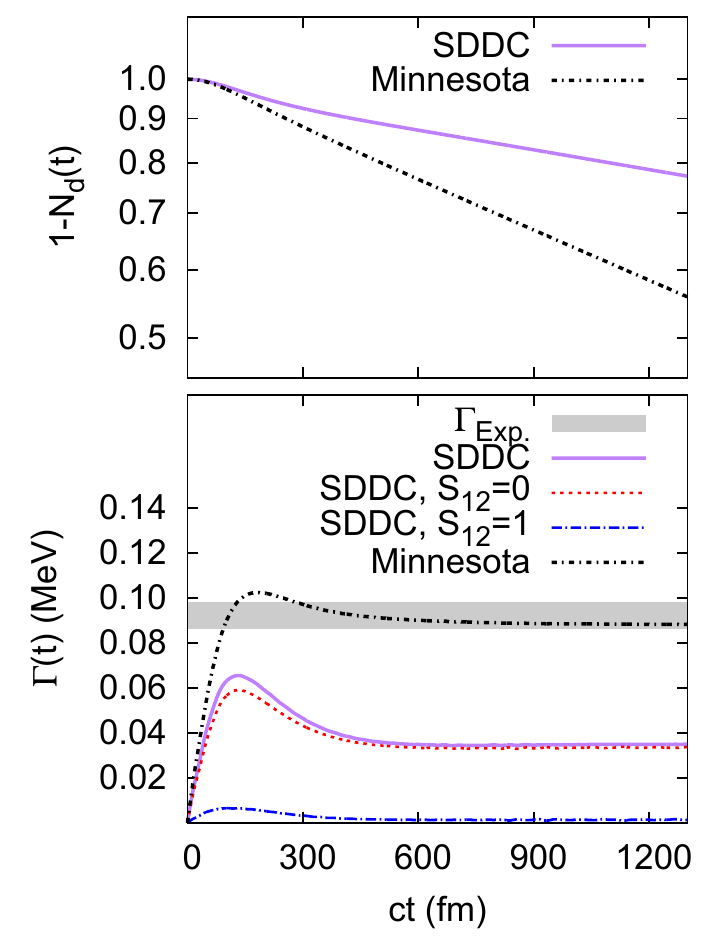}
  \caption{(i) Top panel: survival probability, $1-N_d(t)$, obtained with the surface SDDC pairing interaction. 
For a comparison, the same result but with the finite-range Minnesota pairing 
is taken from Ref. \cite{14Oishi}. 
These are plotted in logarithmic scale. 
(ii) Bottom panel: \twop-decay width of $^6$Be calculated with the 
surface SDDC and the finite-range Minnesota pairing interactions. 
In SDDC case, the spin-singlet ($S_{12}=0$) and spin-triplet ($S_{12}=1$) widths are both plotted. 
Experimental result, $\Gamma_{\rm 2p}=92\pm6$ keV, is indicated by the shaded area \cite{88Ajzen}. 
} \label{fig:plpg}
\end{center} \end{figure}

\begin{table*}[tb] \begin{center}
\caption{Parameters for SDDC pairing interactions 
used in Sec. \ref{Sec:ddd} (upper three rows) and 
Sec. \ref{Sec:asympt} (middle two rows). 
Resultant $Q_{\rm 2p}$ and $\Gamma$ are displayed 
in the same manner as Table \ref{table:CP}. }\label{table:TR}
  \catcode`? = \active \def?{\phantom{0}} 
  \begingroup \renewcommand{\arraystretch}{1.2}
\begin{tabular*}{\hsize} { @{\extracolsep{\fill}} ccccc cc} \hline \hline
              & $w_0$              & $R_f$  & $a_f$        & $\eta$  & $Q_{\rm 2p}$ & $\Gamma$ \\
              & (MeV$\cdot$fm$^3$) & (fm)   & (fm)         &         & (keV)      & (keV)   \\ \hline
~Default SDDC & $-767.398$      & $1.68$   & $0.615$ & $1.04$?  & $1370.7$  & $34.7$   \\
~Steep        & (same)          & $0.84$   & (same) & $2.53$?  & $1370.8$  & $33.3$ \\
~Smooth       & (same)          & $4.50$   & (same) & $0.349$  & $1369.6$  & $33.5$ \\
\hline
~Volume       & ?$-525.5$       & $1.68$   & $0.615$ & $0$     &  $1370.6$   & $19.7$ \\
~Emitter      & $-1036.8$       & (same)   & (same)  & $1.80$?  &  $1367.6$   & $90.3$ \\
\hline
~Experiment \cite{88Ajzen}  &&&&              &  $1371\pm 5$ & $92\pm6 $?? \\
\hline \hline
\end{tabular*}
  \endgroup
  \catcode`? = 12 
\end{center} \end{table*}

\subsection{Decay width} \label{Subsec:width}
We next investigate the decay width, which is one of 
the directly measurable quantities of the \twop~emission. 
From the decay probability, $N_d(t)$, 
the \twop-decay width is calculated as 
\beq
 \Gamma (t) = -\hbar \frac{d}{dt} \ln \left[ 1-N_d(t) \right] = \frac{\hbar }{1-N_d(t)} \frac{d}{dt} N_d(t), \label{eq:width} 
\eeq
where $1-N_d(t)$ indicates the survival probability. 
It is worthwhile to note that, if the time-evolution follows the exponential decay rule, 
which is a fundamental property of radioactive processes, 
the decay probability is given as $N_d(t) = 1-\exp(-t\Gamma_c/\hbar)$. 
Here $\Gamma_c $ is the constant decay width. 
In this case, $\Gamma(t)$ becomes identical to $\Gamma_c$, 
which determines the mean lifetime of this system, $\tau = \hbar / \Gamma_c$. 

\begin{figure}[tb] \begin{center}
  \includegraphics[width = \hsize]{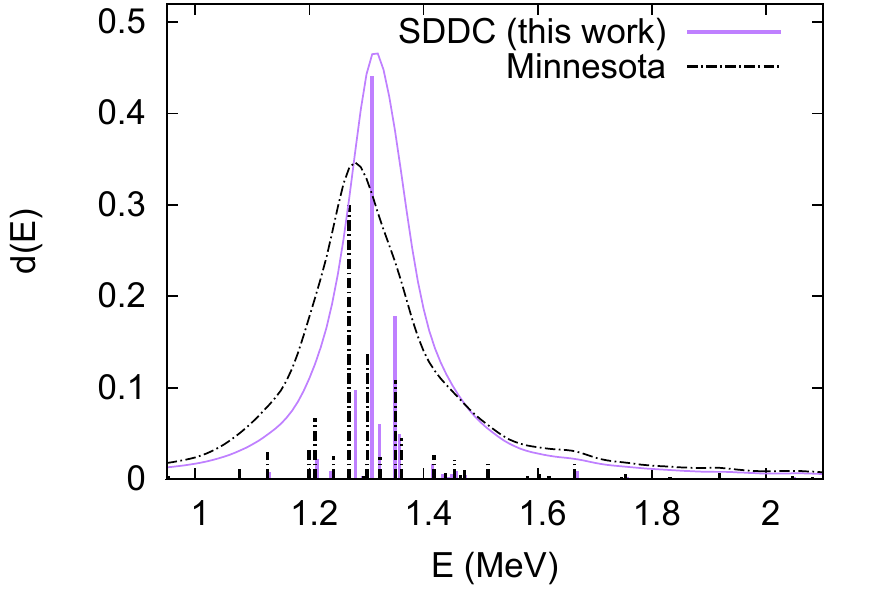} \\
  \caption{Time-invariant discrete energy distribution, $d(E_N)$, obtained 
with the surface SDDC pairing interaction. 
Its continuous distribution, plotted in the arbitrary scale, 
is obtained by smearing $d(E_N)$ with a Cauchy-Lorentz 
function, whose full width at the half maximum is $0.1$ MeV. 
The same plot for the Minnesota pairing case is also displayed. 
} \label{fig:6}
\end{center} \end{figure}

In the following, for a comparison with our SDDC model, 
we fetch the same result with a softened finite-range 
Minnesota pairing model \cite{14Oishi}. 
Notice that, as present in Table \ref{table:CP}, 
both pairing models are adjusted so as to 
reproduce the same $Q$ value. 
This is an important constraint because, 
for radioactive processes governed by the quantum tunneling effect, 
even a small difference in the $Q$ value 
can lead to a large change of the decay 
width \cite{2000Gri_PRL,07Gri_III,12Maru,13Deli,2015Lundmark,2016Kobayashi}. 
However, as an intuitive shortcoming in Ref. \cite{14Oishi}, 
we should also warn that 
fitting Minnesota potential to the $Q$ value leads to 
the inconsistency to the experimental scattering length. 
Also, the core-proton potential and the cutoff parameters 
are common in both cases.

In Fig. \ref{fig:plpg}, we plot the 
survival probability and 
decay width as functions of time. 
After a sufficient time-evolution, 
we finely obtain the exponential decay rule and 
thus the convergence of decay width. 
From Krylov-Fock theorem \cite{47Kry,89Kuku}, 
this exponential decay coincides with that the energy distribution, 
$d(E_N)$, well approximates the Breit-Wigner spectrum. 

For the deviation from exponential decay rule in 
long-time scale, there have been several statements 
of its existence in radioactive 
processes \cite{1961Winter,1977Nico,2002Dicus,2009Ivanov,2014Peshkin}. 
Investigation of this long-time deviation is, however, 
not feasible with present time-dependent model, 
because the reflected wave at $R_{\rm box}$ invokes an unphysical deviation. 
In order to disinfect this ``contamination'' by the unphysically reflected wave, 
one needs to employ, {\it e.g.} absorption boundary condition \cite{1998Brinet,2016Schuet}. 
Because this improvement is technically demanding, we leave it for future work. 
We emphasize that our conclusion based on the resultant decay width 
is independent of this reflected contamination.

In Table \ref{table:CP}, $\Gamma(t)$ value at $ct=1000$ fm is tabulated. 
In our result, 
the SDDC pairing interaction underestimates the experimental \twop-decay width, 
whereas the Minnesota pairing showed a good agreement with it. 
In Fig. \ref{fig:plpg}, the partial decay widths for the spin-singlet and 
spin-triplet channels are also plotted \cite{14Oishi}: 
$\Gamma(t)=\Gamma_{S=0}(t)+\Gamma_{S=1}(t)$. 
One finds again the dominance of the spin-singlet configuration in \twop~emission 
consistently to the density distribution in Fig. \ref{fig:Dens_tide}. 
The exact values of $\Gamma_{S=0,1}(t)$ at $ct=1000$ fm are 
also summarized in Table \ref{table:CP}. 
With the SDDC pairing, the spin-singlet \twop-decay width is 
remarkably small compared with the Minnesota pairing case, 
whereas the spin-triplet width shows the similar values. 
Because of the same setting except the pairing models in two cases, 
the different \twop-decay widths 
should be purely attributed to the pairing properties. 

Figure \ref{fig:6} displays the discrete energy spectra, $d(E_N)$, and their 
continuous distributions smeared by a Cauchy-Lorentz function. 
The spectrum width obtained with the SDDC pairing model 
is narrower than that with the Minnesota pairing model. 
This result coincides with the converged $\Gamma$ values.


Lastly, we confirmed that our conclusion does not change even if 
we employ a smaller value of $E_{\rm cut}$: 
we calculated the decay width with $E_{\rm cut}=32$ MeV, 
with the SDDC pairing interaction refitted to reproduce the 
empirical scattering length and the $Q$ value, $Q_{\rm 2p}=1.37$ MeV. 
Then we obtained the same decay-width value 
as in the original energy cutoff case in Table \ref{table:CP}.

\subsection{Density dependence} \label{Sec:ddd}
In this part, we discuss in detail the inconsistency 
of $\Gamma_{\rm 2p}$ and $Q_{\rm 2p}$. 
First we should remember that the asymptotic pairing strength, $w_0$, 
has been adjusted consistently to the vacuum scattering length, $a_0$, 
which gives the first constraint from experimental results. 
For other two observables, 
$Q_{\rm 2p}$ and $\Gamma_{\rm 2p}$ to be reproduced simultaneously, 
we can manipulate the density-dependence, $w(r)$, 
only around the core nucleus. 
\begin{figure}[tb] \begin{center}
  \includegraphics[width = \hsize]{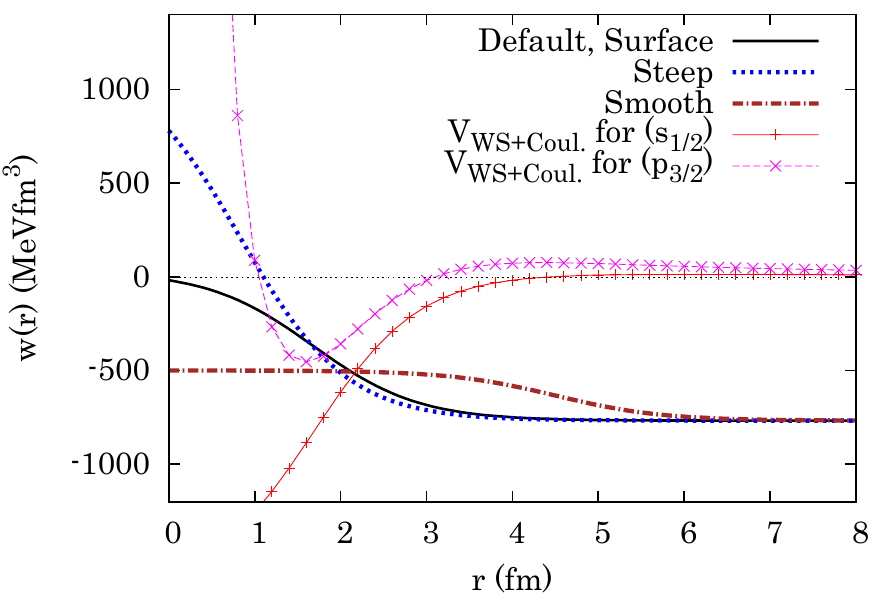} \\ 
  \includegraphics[width = \hsize]{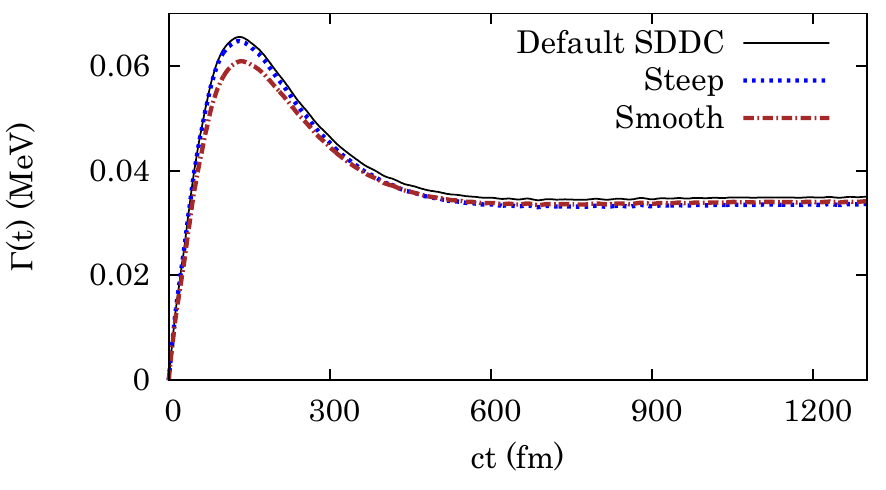} 
  \caption{(i) Top panel: density-dependence of 
the SDDC pairing interaction, $w(r)$, in the default (surface), 
steep and smooth cases. 
The core-proton potential is also plotted in the arbitrary scale. 
(ii) Bottom panel: corresponding result of \twop-decay width. 
} \label{fig:v03}
\end{center} \end{figure}

For this purpose, 
in addition to our default SDDC parameters, 
we test two sets of parameters, {\it steep} and {\it smooth}, 
as summarized in Table \ref{table:TR}. 
Visual plots of these $w(r)$ show in Fig. \ref{fig:v03}. 
In these cases, we modify the radial parameter, 
$R_f$ in Eq.(\ref{eq:w_SDDC}), from the default value. 
Then, we re-adjust the parameter $\eta$ to reproduce the 
empirical $Q$ value. 
Consequently, in the steep SDDC case, the density-dependent strength 
should be positive deeply inside the core, 
meaning that the \twop-interaction should be repulsive there 
due to our $Q$ value fitting purpose. 
On the other hand, in the smooth SDDC case, 
$w(r)$ is always attractive with a smooth change 
around the core-proton barrier. 
We remind that the asymptotic value, $w_0$, is common in all the cases. 
Note also that we change these parameters only 
in the pairing interaction, whereas 
the core-proton interaction, $V_{\rm WS}(r)$, has 
been common in all the cases. 
Namely, resonance parameters of $\alpha-p$ keep unchanged.

In Fig. \ref{fig:v03}, our resultant $\Gamma(t)$ are present: 
there is actually no significant difference in the three cases. 
Namely, the density-dependence of pairing strength plays 
a minor role in the \twop-penetrability, 
whereas only the asymptotic strength can control it. 
It also means that 
there has been no way to resolve the trinity problem of 
$Q_{\rm 2p}$, $\Gamma_{\rm 2p}$, and $a_0$, as long as 
with the simple SDDC pairing model. 
Indeed, this impossible trinity was found also in Ref.\cite{14Oishi}, 
where the softened Minnesota model should 
affect the consistency to the experimental scattering length.

\subsection{Sensitivity to asymptotic interaction} \label{Sec:asympt}
In order to investigate the effect of the asymptotic interaction, 
we repeat the same calculation but changing the 
$w_0$ values in the following. 
Although it leads to an inconsistency to the empirical scattering length, 
we expect to obtain a hint for further 
sophistication of the theoretical model. 
Those sets of parameters are displayed in Table \ref{table:TR}, 
named as {\it volume} and {\it emitter} SDDC interactions. 
In the volume SDDC case, 
we fix $\eta=0$, and fit $w_0$ to the empirical $Q$ value. 
Thus, the pairing strength becomes independent of the radial density. 
This interaction imitates so-called volume type of the pairing energy 
functional in DFT calculations \cite{2005Sand,2008Pastore,2013Pastore}. 
In the emitter SDDC case, on the other hand, 
we search an adequate set of $(\eta,w_0)$, which can 
reproduce the empirical $Q_{\rm 2p}$ and $\Gamma_{\rm 2p}$ simultaneously. 
Consequently, $\eta=1.8$ and $w_0=-1036.8$ MeV$\cdot$fm$^3$ are obtained.

\begin{figure}[tb] \begin{center}
  \includegraphics[width = \hsize]{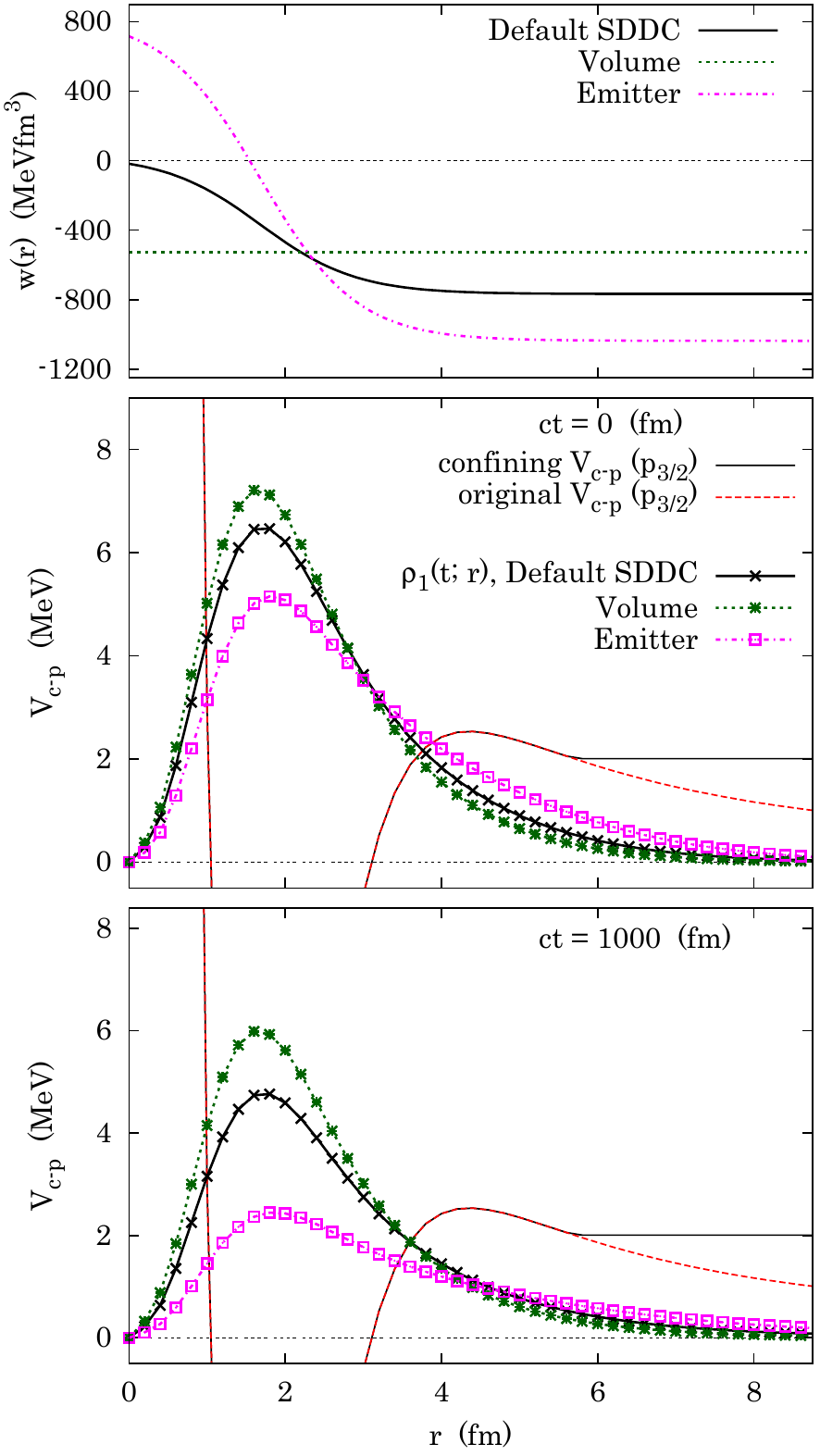}
  \caption{ (i) Top panel: 
The radial strength, $w(r)$, for three SDDC pairing potentials, 
$v_{\rm p-p}^{(N)}(\bir_1,\bir_2)=w(r_1) \cdot \delta(\bir_1-\bir_2)$. 
(ii) Middle and bottom panels: 
The one-proton density-distribution at $ct=0$ and $1000$ fm, respectively. 
These distributions are plotted in the arbitrary scale. 
Confining and original potentials in the $(p_{3/2})$ channel of $\alpha-p$ 
subsystem are also plotted. 
} \label{fig:rho_1p}
\end{center} \end{figure}

In the top panel of Fig. \ref{fig:rho_1p}, 
we plot the contact pairing strength for these SDDC parameterizations. 
It is worthwhile to mention that, with the emitter SDDC model, 
due to its deeper $w_0$ value, 
two protons in vacuum have a larger correlation energy. 

\begin{figure}[tb] \begin{center}
  \includegraphics[width = \hsize]{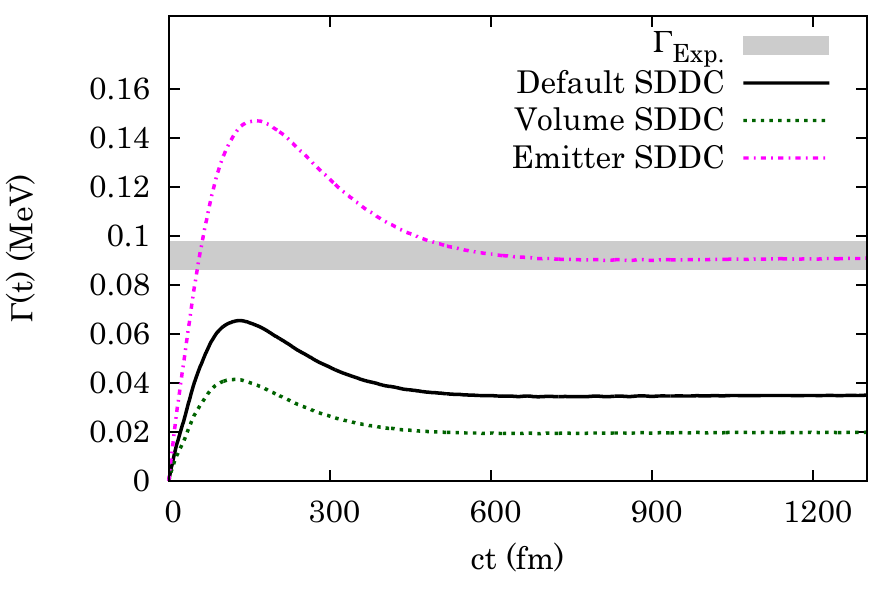} 
  \caption{\twop-decay width of $^6$Be obtained with the default, 
volume, and emitter SDDC pairing models. } \label{fig:plpg52}
\end{center} \end{figure}

Figure \ref{fig:plpg52} shows the decay width obtained with 
different asymptotic strengths. 
Obviously, one can find that the stronger pairing in the 
asymptotic region yields the larger decay width. 
This is consistent to other theoretical 
results \cite{2009Gri_rev,2012Pfu_rev,2000Gri_PRL,07Gri_III,12Maru,13Deli,2015Lundmark}. 
It is also remarkable that 
this asymptotic-pairing sensitivity can be concluded 
even in the equivalent kinematic condition, 
which has been realized with 
the standard $Q$ value in our calculations.

The asymptotic sensitivity may be found 
with other kinds of the pairing model. 
In Appendix, we show another example with 
the Minnesota pairing model, which is not density-dependent 
and has a finite range. 
In that section, by tuning the range and strength parameters of the 
Minnesota pairing, the sensitivity of the \twop-decay width 
is confirmed. 

Our time-dependent model can provide an intuitive way to 
study the asymptotic-pairing sensitivity of \twop-dynamics. 
For this purpose, 
in lower two panels in Fig. \ref{fig:rho_1p}, 
we present the one-proton probability-density distribution of the 
initial and time-developed states. 
That is, 
\beqa
 \rho_1(t;r) &=& 8\pi^2 \int_{0}^{R_{\rm box}} dr_2 r_2^2 \int_{-1}^{1} d(\cos \theta_{12}) \nonumber \\
 && \times \abs{\Psi(t;r,r_2,\theta_{12})}^2. 
\eeqa
Because our \twop-basis functions are anti-symmetrized, 
$\abs{\Psi(\bir_1,\bir_2)}^2$ is symmetric for the 
exchange of $\bir_1$ and $\bir_2$. 
Thus, $\rho_1(t;r)$ represents the mean radial distribution of \twop~probability. 
In these panels, 
with the default (surface) or emitter SDDC, 
the probability density shows a dispersed shape. 
This is of course a product of the strong $p-p$ attraction in vacuum: 
two-proton subsystem more favors the outside from the potential barrier. 
This effect then yields a looser stability corresponding to the larger decay width. 
On the other hand, in the volume SDDC case, two-proton density 
hardly diverges with the isotropically attractive pairing.

In order to resolve the impossible trinity problem in Sec. \ref{Sec:ddd}, 
now the qualitative suggestion appears: 
the pairing model should satisfy both 
(i) the consistency to the asymptotic scattering problem, and 
(ii) the dynamical effect on two protons for tunneling, as seen in Fig. \ref{fig:rho_1p}. 
Possible ways in practice include a non-trivial parameterization of 
the density-dependence \cite{2012Yamagami}, 
and/or the phenomenological three-body force \cite{2009Gri_rev,09Gri_677,01Myo}. 
Note also that, in our present model, this dynamical process 
is driven by the total Hamiltonian, which is not 
time-dependent nor self-consistent to the \twop-state. 
This assumption will need to be concerned 
in forthcoming studies.

\section{Summary} \label{Sec:sum}
We have discussed the dependence of \twop~radioactivity on 
nuclear pairing models within the time-dependent three-body model calculations. 
Comparing the zero-range SDDC and the finite-range Minnesota pairing forces, 
the \twop~dynamics is interpreted as the correlated \twop~emission 
similarly in both cases. 

Evaluating the absolute decay probabilities, 
we found that the two-proton decay width is sensitive to 
the pairing model in usage. 
Utilizing the SDDC parameterizations, 
we showed that the asymptotic strength of the pairing interaction 
essentially controls the \twop-decay width. 
This sensitivity exists even if we exclude the kinematic effect 
by reproducing the equivalent condition on the emitted $Q$ value. 
On the other hand, 
the density-dependence effect around the core plays a 
minor role in this field. 


With the simple SDDC pairing model, 
there remains an impossible trinity problem of $Q_{\rm 2p}, \Gamma$, and 
the two-nucleon scattering length in vacuum, $a_0$. 
In order to reproduce whole of these two-body and three-body 
properties consistently to the experiments, 
further model sophistication is necessary. 
One possible approach is to employ a non-trivial parameterization of 
the density-dependence for the pairing interaction \cite{2012Yamagami}, 
and another is the phenomenological 
three-body force \cite{2009Gri_rev,09Gri_677,01Myo}. 
Because these considerable solutions inevitably harm 
the simplicity of the present model, 
we leave these developments for the future study.

Comparison with other kind of experimental data, 
{\it e.g.} momentum distributions in Refs. \cite{2009Gri_rev,09Gri_677,09Gri_80},
is also an important task for future work. 
For this purpose, however, the present model space should be expanded sufficiently to 
handle with the long-range Coulomb effects. 
Although the computational cost is highly increased, 
it may provides another procedure to validate the pairing models.

Another direction of progress may be the 
implementation of our idea to the meanfield 
calculations \cite{2013Scamp,2014Ebata_pyg,2016Naka_rev,2016Sekizawa}. 
Because our three-body Hamiltonian itself is not time-dependent 
nor self-consistent, 
it is not completely clear whether the similar 
pairing sensitivity exists in the SCMF or DFT calculations. 
Implementing our procedure to 
this framework enables us to perform the systematic 
investigation along the \twop-drip line, 
utilizing the \twop-decay data as the reference quantities. 
A wide experimental survey for the \twop-emitter candidates could be profitable for 
this purpose \cite{13Olsen}.

\begin{acknowledgments}
T. Oishi sincerely thank Lorenzo Fortunato, Andrea Vitturi, 
Kouichi Hagino and Hiroyuki Sagawa for fruitful discussions. 
This work was supported by Academy of Finland and University of Jyvaskyla within 
the FIDIPRO programme and within the Centre of Excellence Programme 2012-2017 
(Nuclear and Accelerator Based Programme at JYFL). 
T. Oishi acknowledge the financial support from 
the PRAT 2015 project {\it IN:Theory} at the University 
of Padova (Project Code: CPDA154713). 
We acknowledge the CSC-IT Center for Science Ltd., Finland, 
for the allocation of computational resources. 
\end{acknowledgments}

\appendix*
\section{Contact and Minnesota Pairing Models}
In this Appendix, we discuss a connection of 
the zero-range pairing model to the Minnesota type, which 
was employed in Ref.\cite{14Oishi}, with several results 
for the \twop-decay width. 
The $p-p$ Minnesota potential used in Ref.\cite{14Oishi} was given as 
\beqa
 v_{\rm p-p}^{(N)} = V_{\rm Min} &=& v_r e^{-d^2/2q^2} \nonumber \\
                           &+& v_s e^{-d^2/2\kappa_s^2 q^2} \cdot \hat{P}_{S=0} \nonumber \\
                           &+& v_t e^{-d^2/2\kappa_t^2 q^2} \cdot \hat{P}_{S=1}, \label{eq:pp_Min}
\eeqa
where $d\equiv \abs{\bir_1-\bir_2}$, 
$v_r=156$ MeV, $v_s=-91.85$ MeV, $v_t=-178$ MeV, 
$q=0.5799$ fm, $\kappa_s=1.788$, and $\kappa_t=1.525$. 
$\hat{P}_{S=0(1)}$ is the projection to the spin-singlet (triplet) channel. 
Remember that $v_r$ was softened from the original value, $v_r=200$ MeV \cite{77Thom}, 
in order to reproduce the reference $Q$ value, $Q_{\rm 2p}=1.37$ MeV. 
Here the first term describes a soft repulsive core. 
\begin{figure}[tb] \begin{center}
 \includegraphics[width = \hsize]{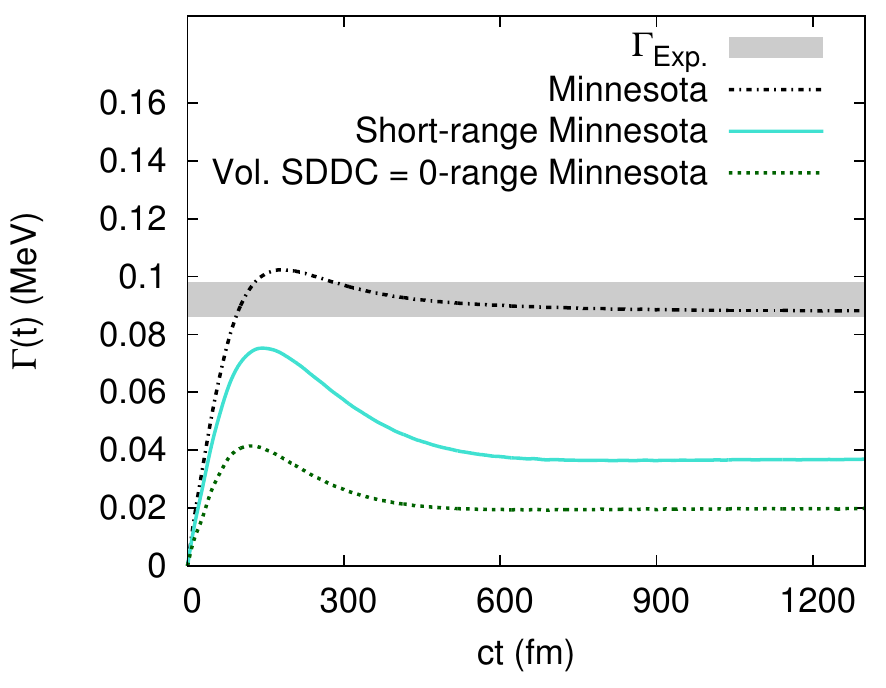}
 \caption{Time-dependent \twop-decay width of $^6$Be obtained with several Minnesota-pairing forces. } \label{fig:10}
\end{center} \end{figure}

Decomposing $d^2=x^2+y^2+z^2$, the first term in Eq. (\ref{eq:pp_Min}) reads 
\beq
 v_r e^{-d^2/2q^2} = w_r \frac{e^{-x^2/2q^2} \cdot e^{-y^2/2q^2} \cdot e^{-z^2/2q^2}}{(2\pi)^{3/2} q^3},
\eeq
where $w_r=v_r(2\pi)^{3/2} q^3=479.1$ MeV$\cdot$fm$^3$, 
and similarly as expected for the following attraction terms. 
Utilizing a well known formula, 
\beq
 \lim_{q\rightarrow 0} \frac{e^{-x^2/2\kappa^2 q^2}}{q\sqrt{2\pi}} = \delta(x/\kappa) = \abs{\kappa} \delta(x), 
\eeq
then at zero-range limit, we obtain 
\beq
 \lim_{q\rightarrow 0} V_{\rm Min}(\bir_1,\bir_2) = w_0 \delta(\bir_1-\bir_2), \label{eq:Minne_ZR}
\eeq
where $w_0=w_r+\kappa_s^3 w_s=-1133.4$ MeV$\cdot$fm$^3$. 
Indeed, this zero-range form is identical to the volume type of 
the pairing force used in Sec. \ref{Sec:asympt}. 
Notice also that, for the two-proton basis in $0^+$ configuration, 
matrix elements of the spin-triplet contact potential become zero 
automatically from the angular-momentum algebra \cite{2012Tani}.

Employing the volume contact pairing given in Eq. (\ref{eq:Minne_ZR}), 
however, we confirmed 
that the $\alpha+p+p$ three-body system fictionally 
becomes bound with $Q_{\rm 2p}\simeq -1.3$ MeV. 
In order to reproduce the experimental $Q$ value, 
we need to use the shallower strength as in Table \ref{table:TR} 
in the main text. 
Then, we obtain $\Gamma=88.2$ and $19.7$ keV with the 
finite-range and zero-range Minnesota potentials, respectively. 
To reinforce our result, we repeat the same calculation but with 
the shorter range, $q/\sqrt{2}\simeq 0.41$ fm in Eq.(\ref{eq:pp_Min}). 
In this case, we need to employ the enhancement factor, $f=2.047$, to reproduce the 
reference $Q$ value: $v_{\rm p-p}^{(N)} = f \cdot V_{\rm Min}(q/\sqrt{2})$.

In Fig.\ref{fig:10}, 
all the resultant \twop-decay widths are displayed. 
As expected, the short-range Minnesota case yields the medium value of the decay width 
between the default and zero-range Minnesota cases. 
Because Minnesota forces are density-independent, 
this sensitivity of \twop-decay width is purely 
attributable to the asymptotic scattering property, 
which is governed by the choice of parameters.


%

\end{document}